\documentclass[aps,prd,groupedaddress]{revtex4}
\usepackage{graphicx}
\usepackage{epsfig}

\def\p0{\phi_{0}}

\def\Sun{\odot}

\def\cns{n_{sim}}
\def\cn{n_x}
\def\t{\tilde}

\def\vecx{{\bf x}}

\def\veck{{\bf k}}

\def\cP{{\cal P}}
\long\def\comment#1{}

\def\bfF{{\bf F}}
\def\bfC{{\bf C}}
\def\hbfC{{\bf \hat C}}
\def\bbfC{{\bf \bar C}}
\def\tbfC{{\bf \tilde C}}
\def\W2{{\cal W}}

\def\vecx{{\mbox{\boldmath $x$}}}

\newcommand{\bi}{B_{l_1 l_2 l_3}}

\def\ben{\begin{enumerate}}
\def\een{\end{enumerate}}
\def\bi{\begin{itemize}}
\def\ei{\end{itemize}}
\def\be{\begin{equation}}
\def\ee{\end{equation}}
\def\bea{\begin{eqnarray}}
\def\eea{\end{eqnarray}}

\def\cmm2{{\,\rm cm^{-2}}}
\def\cm2{{\,{\rm cm}^2}}
\def\cmm3{{\,{\rm cm}^{-3}}}
\def\gcmm3{{\,{\rm g\,cm^{-3}}}}

\def\fun#1#2{\lower3.6pt\vbox{\baselineskip0pt\lineskip.9pt
  \ialign{$\mathsurround=0pt#1\hfil##\hfil$\crcr#2\crcr\sim\crcr}}}

\def\t{\tilde}

\newcommand{\ie}{{\it i.e. }}
\newcommand{\eg}{{\it e.g. }}

\begin{document}

\date{\today}
\title{The Non-Linear Fisher Information content of cosmic shear surveys}
\author{Olivier Dor$\acute{e}$, Tingting Lu, Ue-Li Pen}
\affiliation{CITA, University of Toronto, 60 St George Street, Toronto,
  ON, M5S 3H8, CANADA\\ {\texttt olivier@cita.utoronto.ca, ttlu@cita.utoronto.ca, pen@cita.utoronto.ca}}

\begin{abstract}
We quantify the Fisher information content of the cosmic shear survey two-point
function as a function of noise and resolution. The two point
information of dark matter saturates at the trans-linear scale. We investigate 
the impact of non-linear non-Gaussianity on the information content
for lensing, which probes the same dark matter. To do so we heavily
utilize N-body simulations in order to probe accurately the 
non-linear regime. While we find that even in a perfect survey, there
is no clear saturation scale, we observe that non-linear growth
induced non-Gaussianity could lead to a factor of ~4 reduction for the
common Dark Energy figure of merit. This effect is however mitigated
by realistic levels of shot noise and we find that for future surveys,
the effect is closer to a factor of 1.5. To do so, we develop a new
scheme to compute the relevant covariant matrix. It leads us to claim
an unbiased estimator with an order of magnitude improvement in
accuracy with only twice more simulations than previously
used. Finally, we evaluate the error on the errors using bootstrap methods.
\end{abstract}
\maketitle

\renewcommand{\thefootnote}{\arabic{footnote}}
\setcounter{footnote}{0}

\section{Introduction}

What is the information content of cosmic shear surveys? Although this
question has been addressed many times given the current interest in cosmic shear
surveys, to answer it accurately turns out to be a non-trivial
task. It is the purpose of this work to offer one answer to this question. 

The measure of cosmic shear (\eg \cite{Bartelmann:1999yn,VanWaerbeke:2003uq} for
reviews), is considered to be one of the most promising observational
tools to understand the origin of the accelerating expansion of the
universe \cite{Albrecht:2006um}.  Commonly attributed to the existence
of some extra unknown physics loosely labelled Dark Energy (DE), its
exact nature became one of the salient questions in contemporary
cosmology
\cite{Riess:2004nr,Astier:2005qq,Eisenstein:2005su,Cole:2005sx,Tegmark:2006az,Komatsu:2008hk,Kilbinger:2008gk}. Characterising
the physical properties of DE is the main scientific driver for the
development of new and very ambitious surveys. 

Our work was motivated by recent investigations of the information
content of the 3D matter power spectrum as quantified by the
projection of the Fisher information on the amplitude
\cite{Rimes:2005xs,Rimes:2005dz,Hamilton:2005dx,Neyrinck:2006xd}. Quantifying   
the information through this well defined statistic (insightful even if restrictive), the
answer obtained in these papers was somewhat surprising. While a
Gaussian like behavior was observed on linear (large) scales,  the
non-linear growth of structures entails an information saturation at
\emph{mildly non-linear} scales. A quasi-Gaussian behavior was recovered once
fully in the non-linear regime but at a substantially lower level. When devising a survey, it means that
optimizing the survey to gain sensitivity and resolution in the
trans-linear regime (where the power spectrum is currently interpreted
cosmologically), would not entail much pay-off if one were focusing on
this statistic only. Despite the fact that the effect of the
non-linear growth of structures had been widely studied before these
works, formulating it this way led to this surprising answer,
heuristically understood within the context of the halo model
\cite{Neyrinck:2006xd}. \citet{Neyrinck:2006zi} later showed that if
we project the Fisher information into other parameters, analogous
behavior are then observed. This validates the insightful value of
the amplitude projection. As a consequence, for the sake of
simplicity, we will loosely call information the projection of the
Fisher information onto various subspaces (amplitude or DE
statistics). And although we will define all our statistics precisely, 
their label as information is definitely restrictive. We will look at
them as a way to highlight the departure from the Gaussian behavior
usually assumed when forecasting the constraining power of those
survey. Beyond this departure, to quantify fully the information
content in the non-Gaussian regime is a task we will not endeavor in
this paper.  This paper explores the variance of the power spectrum,
which is a 4-point statistic of the density field, and its variance,
which is an 8-point statistic.
Potentially, higher order estimators, e.g. 3 or 4 points functions,
could contain additional information for non-Gaussian fields.  The
calculation of those errors, and errors on errors, is substantially
more challenging.  Early investigations \cite{Pen:2003vw} indicated
that those errors grow rapidly, making their use challenging.

Whereas new cosmic shear surveys are being advocated
\cite{Albrecht:2006um} and designed \cite{Amara:2006kp}, we want to
study in this work how this 3D information saturation translates into cosmic shear
observables. In particular, one question we would like to answer is
whether there exists a scale above which the Fisher Information for
the two-point functions (projected onto the amplitude or the DE Figure
of Merit) saturates. Since above a given angular scale ($\ell\geq
2000$), our lack of precise modeling of the physics of baryons might require enormous efforts to
be addressed \cite{White:2004kv,Zhan:2004wq,Rudd:2007zx}, it would be
interesting to know whether such a saturation happens and in particular
how it compares to this \emph{theoretical uncertainty} scale. To tackle this question will require to
compute the cosmic shear error budget in the fully non-linear regime. While this
questions has already been investigated in the literature
\cite{White:1999xa,Cooray:2000ry,Semboloni:2006gc,Takada:2008fn}, we
will address it using numerical N-body simulations to probe accurately
the full non-linear regime (still neglecting baryons though), and a new
way to build the covariance matrix from those quantities. This will
lead us to an order improvement in accuracy as compared to previous numerical
works in the literature. We will quantify this statement by measuring
the errors on the errors using bootstrap techniques. 

In this paper, we first begin by introducing the methodology of our
work in Sec.~\ref{sec:metho} before revisiting the 3D matter power
spectrum results in Sec.~\ref{sec:matter_power} as an introduction to 
the cosmic shear case
developed in Sec.~\ref{Sec:CS_inf}. We discuss in Sec.~\ref{sec:disc} the practical
consequences of these results for current and coming optical surveys,
as well as for CMB lensing.

\section{Methodology}
\label{sec:metho}

If we want to infer a set of parameters $\alpha_i$ from observables
$\vecx$ of dimension $\cn$ following a multi-variate Gaussian
distribution with a covariance matrix $\bfC \equiv$$
\langle\vecx\vecx^t \rangle-\langle\vecx \rangle\langle\vecx\rangle$,
the Fisher Information matrix is defined as
\cite{Fisher:1936et,Tegmark:1996bz,Tegmark:1997yq} 
\be
\label{eq:fisher_def}
\bfF_{ij}  \equiv  {1\over 2} {\rm tr}\left[\bfC^{-1}{\partial \bfC \over
    \partial \alpha_i}\bfC^{-1}{\partial \bfC \over \partial \alpha_j}  
  \right] + {\partial \langle\vecx\rangle \over
    \partial \alpha_i}\bfC^{-1}{\partial \langle\vecx\rangle\over \partial \alpha_j}.
\ee
Its relevance for parameter estimation can be seen from the
Cram\'er-Rao inequality stating that the Fisher matrix sets a lower
bound on how well a parameter $\alpha_i$ can be measured, that is
$\sigma^2(\alpha_i^{})\geq 1/\bfF_{ii}^{}$.  
We assume from now on that the covariance matrix $\bfC$ does not depend
on $\alpha_i$ (for a discussion in the context of cosmic shear, see
\cite{Eifler:2008gx}) and, following \citet{Rimes:2005xs}, we define
the information content of $\vecx$ as  
\be
Inf \equiv \sum_{ij} \bfF_{ij} = \sum_{ij}  {\partial \langle\vecx\rangle \over
    \partial \alpha_i}\bfC^{-1}{\partial \langle\vecx\rangle\over \partial \alpha_j}.
\label{eq:inf_dim}
\ee
In this paper, the observables we will consider will be either the 3D
matter power spectra, $\vecx = P(\veck)$, or the 2D convergence
power spectra $\vecx = C_\ell^\kappa$ defined in Eq.~\ref{eq:clkappa_def}. The parameters we will focus
will be the standard cosmological parameters for a flat
cosmological model whose density is dominated at late time by Dark
Energy (DE) whose equation of state evolves as $w=w_0+w_a(1-a)$,
$\alpha=(w_0,w_a,\omega_m,\omega_b,n_s,\sigma_8)$ \cite{Albrecht:2006um}. The nominal
value for those parameters correspond to the currently favored model, $\alpha= (1.,0.,0.1334,0.0228,0.963,0.796)$ \cite{Komatsu:2008hk}.

For pedagogical reasons, we will also consider a dimensionless version
of Eq.~\ref{eq:inf_dim}
\bea
\bar{Inf} \equiv \sum_{ij} \bbfC_{ij}^{-1}&, \quad& \bbfC_{ij} =  {\langle\vecx_i\vecx_j\rangle\over\langle\vecx_i\rangle\langle\vecx_j\rangle} .
\label{eq:inf_nodim}
\eea
This form would be obtained from Eq.~\ref{eq:inf_dim} if we were measuring the amplitude of a template $P(\veck)$ (or
$C_\ell^\kappa$), so that the partial derivatives were unity. Note that because of non-linear effects, this
amplitude does not correspond to $\sigma_8$ or the curvature
perturbation amplitude $A_S$. To put it otherwise, what we will define
as information in Sec.~\ref{sec:matter_power} and \ref{Sec:CS_inf}
corresponds to the variance on the amplitude parameter, $\alpha$, if
the observables were to be modeled as $\vecx = \alpha \bar{\vecx}$. We find this
projection of the Fisher Information matrix onto this space to be a
convenient quantity to visualize the property of this matrix. Given this
definition, since we are interested in quantifying the effects on non-linearities on the information content
of the measured convergence angular power spectrum (or matter power
spectrum), it will be particularly insightful to investigate the
scaling of $\bar{Inf}$ with a cut-off scale $\ell_{max}$ ($k_{max}$), that
is the cumulative information content as function of the smallest
(angular) modes measured. The comparison between the scaling on large scale (low $\ell$ and $k$)
where the convergence (matter) field is expected to be Gaussian to the one in
the non-linear regime (high $\ell$ and $k$) will thus be of particular
relevance. For this purpose, we define in the matter power spectrum
and angular power spectrum case (at wavenumber $k_b$ or multipole $\ell$),
\bea
\bbfC_{k_bk_{b'}<k_{max}}    = {\langle
  P_{k_b}P_{k_{b'}}\rangle\over\langle P_{k_b}\rangle\langle
  P_{k_{b'}}\rangle} &\quad, &\bar{Inf}(k_{max})  =
\sum_{k_b,k_{b'}<k_{max}} \bbfC_{k_bk_{b'}}^{-1} \ , \label{eq:inf_nodimpk}\\
\bbfC_{\ell_1\ell_2<\ell_{max}}    = {\langle
  C_{\ell_1}^{\kappa z_1z_2}C_{\ell_2}^{\kappa z_3z_4} \rangle\over
  \sqrt{\langle C_{\ell_1}^{\kappa  z_1z_1}\rangle\langle C_{\ell_1}^{\kappa  z_2z_2}\rangle\langle C_{\ell_2}^{\kappa z_3z_3}\rangle\langle C_{\ell_2}^{\kappa z_4z_4}\rangle}} &\quad, &\bar{Inf}(\ell_{max})  =  \sum_{\ell_1,\ell_2<\ell_{max}}
\bbfC_{\ell_1\ell_2}^{-1}\ .\label{eq:inf_nodimcl}
\eea
The definition of $\bar{Inf}$ in Eq.~\ref{eq:inf_nodim} is
particularly easy to interpret, since in the Gaussian case, where $\bfC_{ij}^{}\propto$$
\vecx_i^{}\vecx_j^{}\delta_{ij}^{}$, it directly reduces to half the
    number of measured modes. We thus have a simple analytical
    predictions for the expected scaling on large scales. Note that we choose to
    define $\bar{Inf}(\ell_{max})$ by imposing a sharp
    cut-off in Fourier space. An alternative definition consists in
    marginalizing over all the modes above $k_{max}$ ($\ell_{max}$) by
    adding a white noise level so that for example the signal to
    noise ratio equals 1 at $k=k_{max}$ ($\ell=$$\ell_{max}$). We
    checked that both approaches give equivalent results. 

As visible from Eq.~\ref{eq:inf_nodim}, the key quantities to evaluate
the information content of our observables is the covariance
matrix. For this purpose, we will use a Monte-Carlo approach and generate
$\cns$ realizations of the observables, $\vecx_k$, through N-body
simulations including dark matter only. We make use of the publicly available \texttt{CubePM}
code \footnote{\texttt{http://www.cita.utoronto.ca/$\sim$merz/cubepm/}}. \texttt{CubePM}
is the successor to the particle-mesh N-body code \texttt{PMFAST} \cite{Merz:2004uq}\footnote{\texttt{http://www.cita.utoronto.ca/$\sim$merz/pmfast/}}
. In addition to the features provided by
\texttt{PMFAST} -- support for distributed memory systems through MPI
and shared memory via OpenMP, minimal memory overhead and
communications requirements -- \texttt{CubePM} contains support for
gas evolution through use of a TVD MHD module, particle-particle
interactions at sub grid cell distances, optimal scaling up to (and
hopefully beyond) 1000's of nodes, as well as shared-memory
parallelization via OpenMP to optimize memory usage on shared memory
nodes. 

Given a set of $\cns$ realization for $\vecx_k$ that we write, $\vecx_k^{s=1,\ldots\cns}$, we define as an
estimator for $\bfC$ (see \cite{Hamilton:2005dx} for a thorough
  discussion on how to measure $\bfC$ from one simulation only)
\bea
\tbfC_{ij} & = & {1\over \cns}\sum_{s=1}^{\cns} \left(\vecx_i^s-\mu_i\right)\left(\vecx_j^s-\mu_j\right) , \\
\mu_i & = & {1\over \cns}\sum_{s=1}^{\cns} \vecx_i^s .
\eea
At this point, it is often missed that the inverse of a
maximum-likelihood estimator for a variable $X$ is in general not an
unbiased estimator of the inverse $X^{-1}_{}$
\cite{Hartlap:2006kj}. To remedy this fact, a corrective factor is
required. Since we also evaluate $\mu$ from our simulation, it can be shown that the following
estimator for $\bfC^{-1}$ is unbiased:
\bea
\hbfC^{-1}_{} & = & {\cns-\cn-2\over \cns-1} \tbfC^{-1}_{}.\label{eq:cinv_est}
\eea
In the case we are interested in here, \ie the the convergence
angular (cross-) power spectrum covariance matrix $\langle
C_{\ell_1}^{\kappa z_1z_2}C_{\ell_2}^{\kappa z_3z_4} \rangle$, we found
that the number of independent modes, $\cn$, is not easy to
define. Thus we dropped this corrective factor. We 
carefully checked the convergence of our results by increasing $\cns$
(see Fig.~\ref{fig:inf_matter_power} and Fig.~\ref{fig:inf_cl_kappa_conv}).

Furthermore, to quantify the error on our statements, we will
evaluate the errors on $\hbfC^{-1}_{}$ by making use of
the bootstrap method \cite{Efron:1993}. We will consider 1000 sets of
$\cns$ simulations randomly drawn from our $\cns$ simulations and apply the
above defined formalism to each. This procedure weighs in a random
manner our initial set of N-body simulations. Even though it is
unclear wether the number of independant realizations,
i.e. simulations, we have is enough for the bootstrap method to be
reliable it still gives us a  valuable glimpse at the reliability of
our statements, that is on the error on the error. 

\section{Matter power spectrum information content}
\label{sec:matter_power}

\begin{figure}[tbp]
  \begin{center}
  \epsfig{file=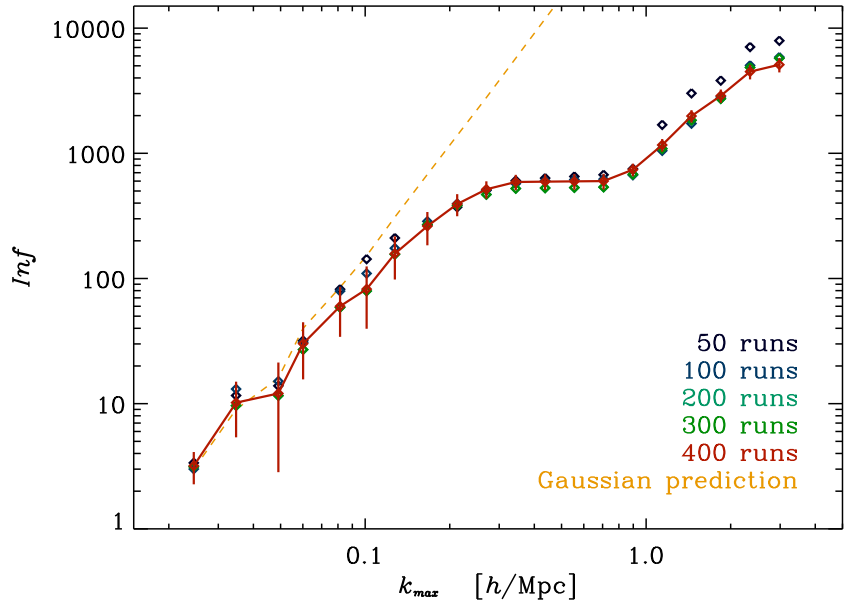,width=0.49\textwidth,angle=90}
  \caption{\footnotesize Cumulative information for the matter power spectrum at $z=1$ as
    defined in Eq.~\ref{eq:inf_nodimpk}. We use either $\cns=$ 50, 100, 200,
    300 or 400 simulations. The bootstrap error bars are obtained
    using an analysis of 1000 sets of 400 simulations. The dashed
    orange line corresponds to the Gaussian prediction, \ie the number of $k$
    modes present in the simulation below $k_{max}$. Although the
    convergence does not appear to be perfect, the difference between the measurement
    using 300 or 400 appear smaller than the error bars everywhere. As
    such we can trust the results obtained with 400 simulations. This
    plot reproduces the results of \citet{Rimes:2005dz}.}
    \label{fig:inf_matter_power}
  \end{center}
\end{figure} 

We first focus on the matter power spectrum and revisit the results of
\citet{Rimes:2005dz}. For this purpose, we ran 400 N-body simulations
with their choice of cosmological model, that is a flat $\Lambda$CDM
model with $\Omega_m=0.29$, $\Omega_{\Lambda}$=0.71, $\Omega_b=0.046$, $\sigma_8=0.97$ and
$h$=0.71. Since we are interested in trans-linear scales, \ie the
transition regime between the fully linear regime and the fully
non-linear regime, a comoving box size of 256 Mpc/$h$ with $256^3$ grid
points is appropriate. This gives us roughly a mass resolution of
9.2$\times 10^{11}M_\Sun$ and a force resolution of 1 Mpc/$h$. The initial conditions were generated at $z=200$.

To improve convergence, to increase the rank of $\tbfC$ and to
alleviate numerical issues when performing the inversion, we define
$\cn =20$ bins logarithmically spaced in $k$ space and we measure the
average power spectrum within a $k$ bin, $\vecx =P_b$ as   
\bea
{(2\pi)^3\over V_{box}} \langle \delta_\veck^{}\delta_\veck^*
\rangle_{\veck \in b}^{}  = k_b^3P_b^{} \ .
\eea
where $\delta_k$ is the Fourier transform of the matter
over-density. We then make use of Eq.~\ref{eq:inf_nodimpk}
and \ref{eq:cinv_est} with $\cns=50,100$, $200$, $300$ or $400$ and $\cn=20$ to compute the
cumulative information content for various $k_{max}$.
The results at $z=1$ are plotted in Fig.~\ref{fig:inf_matter_power} as well as the Gaussian prediction
using the measured number of modes in the simulation,
$\bar{Inf}^{G}_{}(k_{max}^{}) =$$\sum_{b<k_{max}^{}}(\#k\in
b)$$\propto k_{max}^3$. Qualitatively similar results are obtained at different $z$.

First, we notice that the convergence in terms of the number of
simulation used to compute $\bar{Inf}$ seems satisfying although not
perfect. We estimate that the lack of convergence adds
extra-uncertainties of the same order as the error we
estimated using bootstraps. As discovered by \citet{Rimes:2005dz}, the
two remarkable features of this cumulative information are the following. Whereas on linear scales, say $k<$ 0.1
Mpc/$h$ the information content follows the scaling expected from a
Gaussian random field, a sharp transition to a plateau is observed on trans-linear scales,
and a return to a quasi-Gaussian scaling (but with a lower amplitude) in the fully non-linear
regime. These features mean that the information, say on a primordial
amplitude is conserved and could be measured with an accuracy
directly proportional to (\# modes)$^{-1/2}$ on linear scales, not much more is
learned on trans-linear scales, \ie the information is redundant with
the one contained in linear scales. On the other hand, a
quasi-Gaussian scaling is reestablished in the fully non-linear
regime. The sharp transition from the linear to the non-linear regime
can qualitatively be understood in the halo model framework, where 
it corresponds to the transition from the 2 halos term to the 1 halo
term \cite{Neyrinck:2006xd}. Whereas on large scales, the information
is contained in the 2 halos term and scales as the number of modes
measured (or the number of halos in a given volume), on trans-linear
scales, the 1 halo term starts to dominates but with a large variance since most
of its contribution comes from rare massive halos. This large variance
explains why it is hard to extract any information supplemental to the
one obtained in the linear regime from this regime. On smaller
scales though, the 1 halo term contribution comes mostly from numerous
smaller mass halos whose number is much more constant, \ie fluctuates
with much less variance, and the information scales again roughly with the number of modes probed. 
 
Now that we have reproduced and introduced the key results regarding
the 3D matter power spectrum we move to the original results of this paper, that is how this
information saturation effect in the matter power spectrum translates
into some integral of it, \ie into the cosmic shear observables.

\section{Cosmic shear information content}
\label{Sec:CS_inf}

We now make use of $\cns=300$ N-body simulations run with our nominal
cosmology to investigate the information content of cosmic shear surveys. To
quantify the information, we choose as our observable the convergence
cross-power spectra between two redshift bins $z_i$ and $z_j$ that we
defined as \cite{Bartelmann:1999yn,VanWaerbeke:2003uq}
\bea
\label{eq:clkappa_def}
\tilde n_{z_i}^{}\tilde n_{z_j}^{}C_{\kappa \ell}^{z_iz_j}=\int_0^{\infty}
dz\ W^{z_i}_{}(z)W^{z_j}_{}
(z) {H(z)\over D^2_{}(z)}P(\ell/D(z),z)\ ,
\eea
where $H(z)$ is the Hubble parameter, $D(z)$ is the angular diameter
distance,  $P(k,z)$ is the 3-dimensional matter power spectrum at
redshift $z$. The lensing kernel is defined as 
\bea
W^{z_i}_{}(z) & = & {3\over 2}\Omega_m^{}{H_0^2D(z)\over
  H(z)}(1+z)\int_z^{\infty}dz'\  n_{z_j}^{}(z'){D_{LS}^{}(z,z')\over D(z')} \ , \\
\tilde n_{z_i}^{} & = &\int_0^{\infty}dz\ n_{z_i}^{}(z)\ .
\eea
where $D_{LS}^{}(z,z')$ is the angular diameter distance between  $z$
and $z'$, $n_{z_i}^{}(z)$ is the galaxy distribution in redshift bin
$i$ and $\tilde n_{z_i}^{}$ is the total number of galaxies in this
redshift bin. 

In this section, for the sake of simplicity, we will consider a
uniform galaxy distribution in $n_z=1,\ldots, 4$ redshift bins of width $\Delta
z =0.5$ and defined as $1.0<z<1.5$, $1.5<z<2.0$, $2.0<z<2.5$ and
$2.5<z<3.0$. Each distribution is normalized to unity, \ie
$n_{z_i}^{}=1/\Delta z$ and $\tilde n_{z_i}^{}=1$. This choice is
motivated by our interest in low redshift diffuse lensing of the
diffuse radiation originating from the 21cm line emission of
galaxies \cite{Lu:2007pk,Pen:2008fw,Lu:2009}. We will consider more realistic galaxy
distribution functions when discussing specific surveys in
Sec.~\ref{sec:disc}. As discussed in a companion paper \cite{Lu:2009}, we
found that simulations with a box size of 200 Mpc/$h$, a 1024$^3$ grid,
and 512$^3$ particles are close to optimal for our needs. Each one of
these simulations takes about 4.5 hours using 8 nodes (64 cores) on
CITA's Sunnyvale cluster. We checked that finite resolution effects do
not affect the convergence power spectra up to $\ell\simeq 10000$ that
will define the smaller angular scale we consider in this work. This
box size corresponds roughly to an area of 56 square degrees. As a
consequence, when considering the dimensionless cumulative
information, the sum is performed for $\ell\geq 50$.

To compute the covariance matrix using $\cns=300$ simulations we employ an
original method that avoid the artifacts present in previous
methods. The now standard approach to simulate cosmic shear has been
pioneered in \cite{White:1999xa,Jain:1999ir}. It consists in
ray-tracing through a light cone build out of a collection of N-body 
simulations outputs at various redshifts. This method has been widely tested and its limitations
(angular resolution, periodicity, mass resolutions, etc.) quantified
\cite{White:1999xa,Jain:1999ir,Hamana:2000wb,Hamana:2001vz,Semboloni:2006gc}. It
provides great reliability, \eg to produce $\kappa$ maps in the
observational regime of interest nowadays. However, it is important to
notice that given our box length, from the observer at $z=0$ till the
most distant lens plane at $z=3$, 24 boxes are required. As such, in
principle, one could build only 12 fully independent light cones out
of our $\cns=300$, a number far from enough for our Monte-Carlo
approach to compute $\bfC^{-1}$. A common fix consists in using the
same simulations more than once in a given light cone after random
translations and rotations of the original box. While it does increase
the number of light cone realisations that can be generated with a given number
of N-body simulations, it introduces spurious correlations --
density field from a same simulation at different redshift are not
independent -- which are hard to control safely. In fact, from the
Limber approximation we know that the combination of shifting, stacking and
recycling will lead to the correct power spectra since it is a linear
function of the density field at each redshift with random phases. 
But this does not hold
anymore when considering covariance matrices. To remedy this problem,
we follow an original approach described below. 

The basic idea goes as follows. We first compute the covariance matrix
of the convergence (cross-) power spectra for each output boxes
combination by averaging over all the $\cns$ sims. The final
convergence matrix is then an appropriately weighted sum of the
covariance matrices computed at each output redshift. More formally,
this procedure can be written this way. To compute the convergence
power spectra, for each output boxes at a given redshift 
$z_{s_1}$, we project on a randomly chosen side the over-density
field, $\delta(\vecx,z_{s_1})$, Fourier transform it and measure its
2D power spectrum, $\t\delta_{2d}(\veck,z_{s_1})$. After converting
the comoving wavenumber $k$ to an angular multipole
$\ell=kD(z_{s_1})$, we weight the power spectrum by the lensing kernel
$W^{z_{s1}}$ to transform it into an angular convergence power spectra
at  $z_{s_1}$ and bin it in $n_\ell=12$ bins: 
\be
\hat C_\ell^{{z_{s1}},{z_{s2}}} = \langle
\t\kappa^{z_{s1}}_{\ell'}{\t\kappa}^{z_{s2}\,\star}_{\ell'}\rangle^{}_{\ell\pm\Delta \ell} \ ,
\ee
where the average is taken over the $N_\ell$ contributing to this band
power, $\ell-\Delta\ell< \ell'=kD(z_{s_1}) < \ell+\Delta\ell$, where
\be
{\t \kappa}^{z_{s1}}_\ell=\sum_{i}\t\delta_{2d}(\ell=kD(z_{s_1}),z_i) W^{z_{s1}}(z_i) \ ,
\ee
and where $W^{z_{s1}}$ is the lensing kernel for each slices if the sources
are distributed in band $z_{s1}$ and the sum over $i$ denotes a sum
over outputted simulation boxes. It follows that to compute the convergence cross power spectra
covariance matrix,
\bea
Cov\left(C_\ell^{{z_{s1}},{z_{s2}}},C_{\ell'}^{{z_{s3}},{z_{s4}}}\right) &=&
\langle{C_\ell^{{z_{s1}},{z_{s2}}}C_{\ell'}^{{z_{s3}},{z_{s4}}} }\rangle-\langle{C_\ell^{{z_{s1}},{z_{s2}}}}\rangle\langle{C_{\ell'}^{{z_{s3}},{z_{s4}}}
}\rangle\quad ,
\eea
we need to compute using our $\cns$ simulations, both
\bea
\langle C_\ell^{{z_{s1}},{z_{s2}}}C_{\ell'}^{{z_{s3}},{z_{s4}} \star}
\rangle_{sim}^{} &=& \langle{ \t\kappa_\ell^{z_{s1}}
  {\t\kappa}_\ell^{z_{s2}\,\star}   (\t\kappa_{\ell'}^{z_{s3}}
  {\t\kappa}_{\ell'}^{z_{s4}\,\star})^{\star} }\rangle_{sim}^{} \\ 
&=& \sum_{i_1,i_2,i_3,i_4} W^{z_{s1}}(z_{i_1})
W^{z_{s2}}(z_{i_2})W^{z_{s3}}(z_{i_3}) W^{z_{s4}}(z_{i_4}) \langle{
  \t\delta_{2d}(\ell,z_{i_1}) {\t\delta_{2d}}^{\star}(\ell,z_{i_2})
  \t\delta_{2d}^{\star}(\ell',z_{i_3}) {\t\delta_{2d}}(\ell',z_{i_4})
}\rangle_{sim}^{} \nonumber\\  
\eea
and
\bea
\langle{ C_\ell^{{z_{s1}},{z_{s2}}}
}\rangle_{sim}^{}\langle{C_{\ell'}^{{z_{s3}},{z_{s4}}} }\rangle_{sim}^{}&=&\langle{\t\kappa_\ell^{z_{s1}} {\t\kappa}_\ell^{z_{s2}\,\star} }\rangle_{sim}^{}\langle{\t\kappa_{\ell'}^{z_{s3}} {\t\kappa}_{\ell'}^{z_{s4}\,\star} }\rangle_{sim}^{} \\ 
&=& \left[\sum_{i_1} W^{z_{s1}}(z_{i_1})
  W^{z_{s2}}(z_{i_1})\langle{\t\delta_{2d}(\ell,z_{i_1}){\t\delta_{2d}}^{\star}(\ell,z_{i_1})
  }\rangle_{sim}^{}\right] \nonumber\\
&\times& \left[\sum_{i_2} W^{z_{s3}}(z_{i_2})
  W^{z_{s4}}(z_{i_2})\langle{\t\delta_{2d}(\ell',z_{i_2}){\t\delta_{2d}}^{\star}(\ell',z_{i_2})
  }\rangle_{sim}^{}\right]\ .
\eea
Note that care has to be taken regarding the complex
conjugates. Whereas the expectation value for the cross-power spectra
are real, the estimator of the cross power spectra are complex.
\begin{figure}[t]
  \begin{center}
  \epsfig{angle=90,file=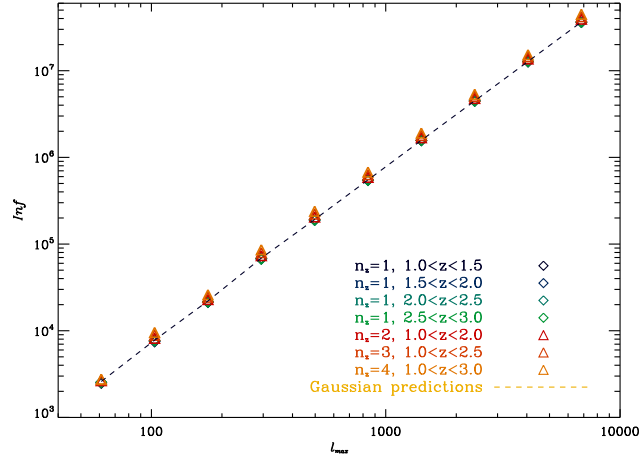,width=0.49\textwidth}
  \caption{\footnotesize Dimensionless cumulative information for the convergence
    cross-power spectra defined in Eq.~\ref{eq:inf_nodimcl}. We here
    replaced the projected density field in each output box by a
    Gaussian field with the same power spectrum as the one measured in
    the N-body box. Each color corresponds to a different sub-set of
    redshift source bands of size 1, 2, 3 and 4. The dashed line
    corresponds to the matching Gaussian predictions using the number
    of modes measured in the simulations.}
    \label{fig:inf_cl_kappa_gauss}
  \end{center}
\end{figure} 

Note that if $\ell\ne \ell'$, the cross terms
$\langle{\t\delta_{2d}^{}(\ell,z){\t\delta_{2d}}^{\star}(\ell',z)}\rangle_{sim}^{}$
do not contribute. After some simple arithmetic, the calculation simplifies to 
\bea
Cov(C_\ell^{{z_{s1}},{z_{s2}}},C_{\ell'}^{{z_{s3}},{z_{s4}}})
&=& \sum_{i_1} W^{z_{s1}}(z_{i_1}) W^{z_{s2}}(z_{i_1}) W^{z_{s3}}(z_{i_1}) W^{z_{s4}}(z_{i_1})
\left[ \langle {\t\delta_{2d}^{}}(\ell,z_{i_1}){\t\delta_{2d}^{\star}}(\ell,z_{i_1}){\t\delta_{2d}^{}}(\ell',z_{i_1}){\t\delta_{2d}^{\star}}(\ell',z_{i_1})\rangle_{sim}^{}\right.  \nonumber \\
&-&\left. \langle \t\delta_{2d}(\ell,z_{i_1})\t\delta_{2d}^{\star}(\ell,z_{i_1})\rangle_{sim}^{}\langle \t\delta_{2d}(\ell',z_{i_1})\t\delta_{2d}^{\star}(\ell',z_{i_1})\rangle_{sim}^{}\right] \nonumber \\
&+&\sum_{i_1,i_2,i_1\ne i_2} W^{z_{s1}}(z_{i_1}) W^{z_{s2}}(z_{i_2})W^{z_{s3}}(z_{i_1}) W^{z_{s4}}(z_{i_2}) \langle \t\delta_{2d}(\ell,z_{i_1})\t\delta_{2d}^{\star}(\ell',z_{i_1})\rangle_{sim}^{} \langle \t\delta_{2d}(\ell',z_{i_2})\t\delta_{2d}^{\star}(\ell,z_{i_2})\rangle_{sim}^{} \nonumber\\
\eea
For the sake of clarity, we omitted the sum over $\ell$  and
$\ell'$ modes within a band power. As compared to the common approach that consists in building light
cones, from which kappa maps and the associated angular (cross-) power
spectra are build, the advantages of our method are two-folds. First, we ensure that
there is no contamination due to the recycling of boxes from the same
simulation. While it has been tested that this contamination is a small effect at
the power spectrum level \cite{Hamana:2000wb}, it does introduce
biases when computing the four point functions of interest to
us; those biases have not been properly quantified yet. In fact, as
was said earlier, the Limber approximation guaranties the former, it
also is generically biased.  A upward fluctuation in the initial condition
will lead to correlated upward fluctuation at all redshifts.
In our scheme, cross terms between redshifts are explicitly not present.

\begin{figure}[t]
  \begin{center}
  \epsfig{angle=90,file=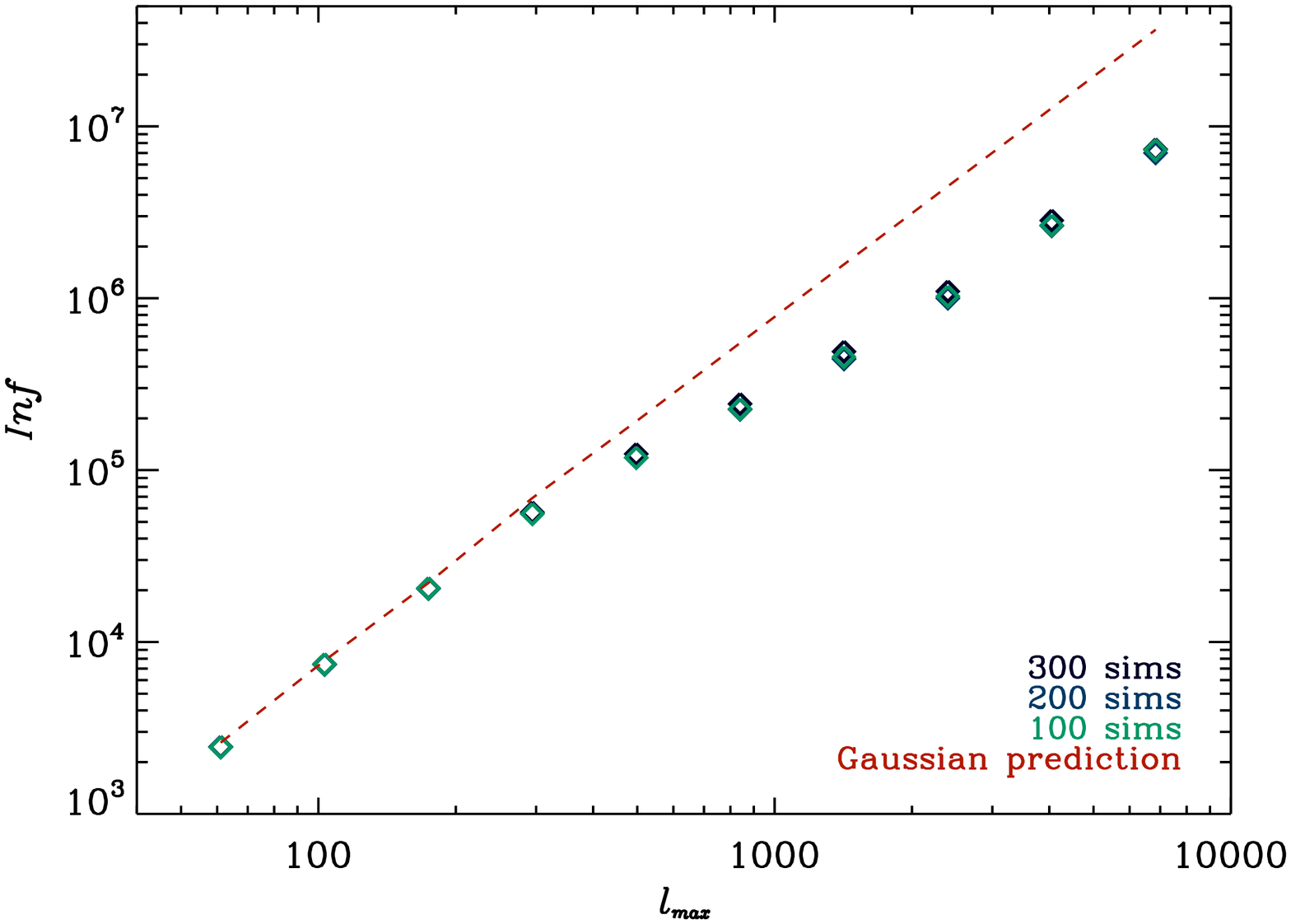,width=0.49\textwidth}
  \epsfig{angle=90,file=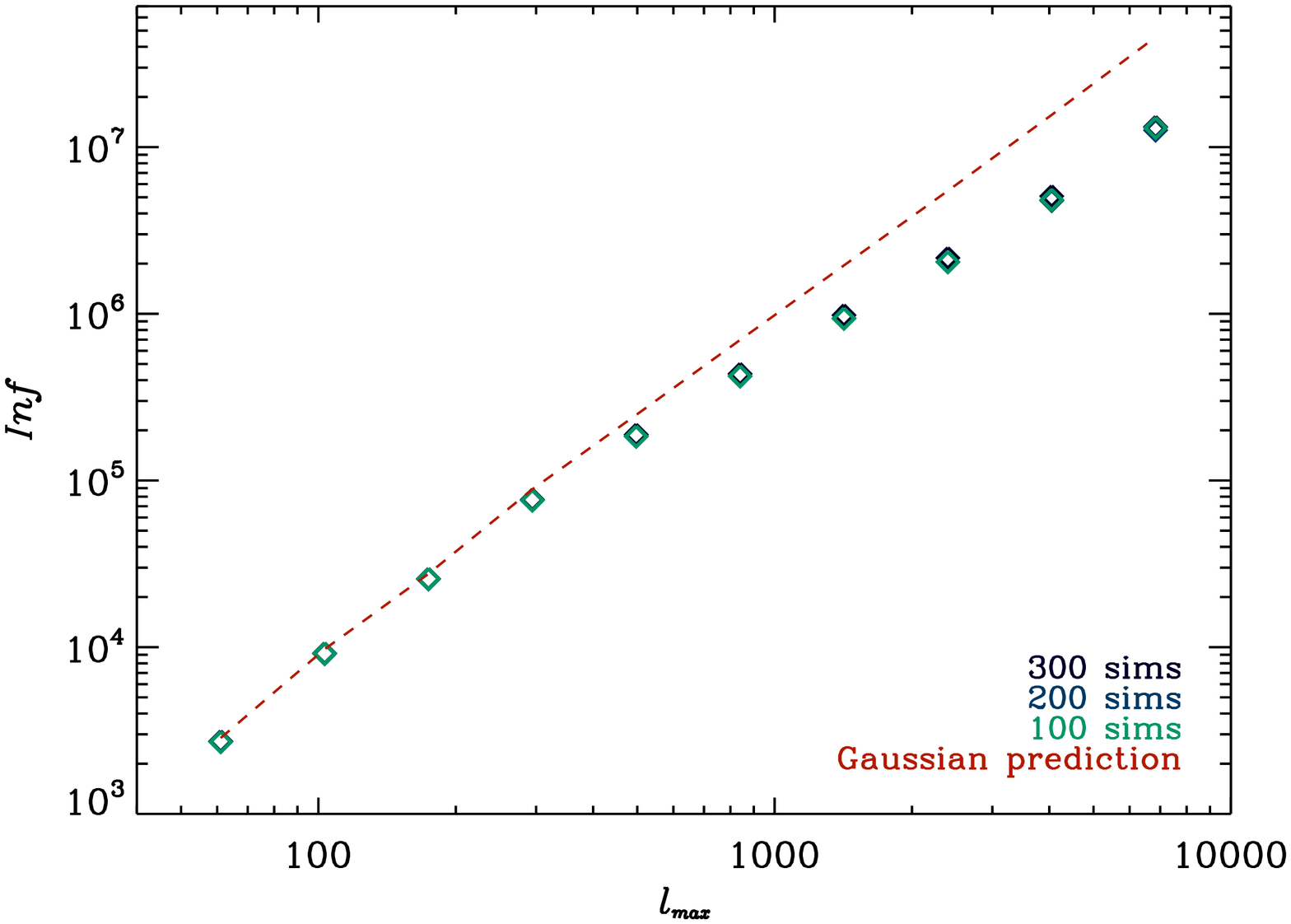,width=0.49\textwidth}
  \caption{\footnotesize Dimensionless cumulative information for the convergence
    cross-power spectra defined in Eq.~\ref{eq:inf_nodimcl} when considering either one
    redshift source bin (1$<z<$1.5, left plot) or four (1$<z<$1.5, \ldots 2.5$<z<$3.0, right plot). Each color
    corresponds to a different number of simulations (100, 200 or 300)
    used to compute the covariance matrix. The weak scatter amongst
    those different points allows us to assess the satisfying level of
    convergence we obtain with 300 simulations.}
    \label{fig:inf_cl_kappa_conv}
  \end{center}
\end{figure}

Similarly, while assuring an unbiased Fisher matrix, this scheme
could still result in a bias on its error,
since the output of the
same $\cns$ simulations is used to compute the covariance matrix of the convergence power spectra at each simulation
output redshift. Second, the rate of convergence with the number of
simulation is much faster than with the usual light-cone
approach. Imagine we need $n_{box}$ to build a full light cone
($n_{box}=24$ in our case). The number of independent light cone is
thus $\cns/n_{box}$ and the rate of convergence for the covariance
matrix goes as $\sqrt{n_{box}/\cns}$. In our case, the convergence rate for
each covariance matrix at each output redshift goes as
$1/\sqrt{\cns}$. Furthermore, on small angular scales, because the
non-linear evolution makes the (high) $k$ evolution quite independent
from one redshift to another, we gain an additional factor $1/\sqrt{n_{box}}$
so that the convergence is closer to $1/\sqrt{\cns n_{box}}$. This effect
is less important on large angular scales (linear) however, on all
scales, the $1/\sqrt{n_{box}}$ scaling will still be regained because various
$k$ modes will appropriated to various $\ell$ bin so that when the
errors will still average down as $1/\sqrt{n_{box}}$ when computing the final
covariance matrix. We thus claim a convergence improvement close to
$1/n_{box}$ ($= 24$ in our case) as compared to the standard
method, \ie one order of magnitude improvement (see
Fig.~\ref{fig:inf_cl_kappa_conv} and discussion in
Sec.~\ref{sec:disc}). 

For cross-checking purposes, we compared all our results with the
usual light cone method. We observe a satisfying agreement between
both. We also clearly observe the different convergence speed as $\cns$ grows.   

Once the $\cn = n_\ell n_{z_{bin}}$$(n_{z_{bin}}+1)/2$ dimensional
covariance matrix has been computed, we apply Eq.~\ref{eq:cinv_est} to obtain an estimate
$\hbfC^{-1}$. As a first check, we consider a set of Gaussian
simulations where the 2D projected density field at each output redshift is replaced by a realisation of a
Gaussian field with an identical spectrum as the one resulting from
N-body simulations. In that particular case, given the exact number of
modes in each $\ell$ shell, we can predict exactly the scaling of the
dimensionless cumulative information and compare it with the
measurements. We consider different number of source bins and all
the associated angular cross-power spectra. The results are displayed in
Fig.~\ref{fig:inf_cl_kappa_gauss} and an excellent agreement is observed
between the analytical predictions (dashed line) and the measured quantities (symbols).

We then move to the genuine N-body simulations and look at the
dimensionless information for either one redshift source bin (left panel of
Fig.~\ref{fig:inf_cl_kappa_conv}) or four redshift source bins (right panel of
Fig.~\ref{fig:inf_cl_kappa_conv}). In both plots we check the
convergence of our result by varying the number of used
simulations and compare it with the Gaussian predictions (red dashed line). We considered respectively 100, 200 and 300
simulations. Obviously the convergence is satisfying and much smaller
than the effects we are interested in, \ie the difference between the
red dashed line and the symbols. As discussed in \cite{Lu:2009}, 
when assessing the convergence of a Monte-Carlo
estimator of $\hbfC^{-1}_{}$ is it important to look at $\hbfC^{-1}_{}$ (or its
norm) and not the diagonal of $\hbfC$ as in
\cite{Takahashi:2009bq}. The relative difference between the 200 and
300 simulations computation, probably an upper bound on the
convergence, is consistent with our convergence estimate. It is around 8\%.

This satisfying convergence is crucial and gives us confidence in the
results displayed in Fig.~\ref{fig:inf_cl_kappa_conv}. We can now
interpret them. First, it is interesting that the saturation effect
present in the 3D power spectrum (see Fig.~\ref{fig:inf_matter_power})
also appears in the  convergence power spectrum. As expected, whereas
N-body results agree with Gaussian predictions on large (linear) 
angular scales, a departure from the Gaussian behaviour (red dashed
line) appears at sub-degree scales ($\ell>300$). Not surprisingly, for sources between
($1<z<1.5$), this corresponds to $k\simeq 0.2 h/$Mpc at $z=0.5$ where
the lensing kernel peaks, consistent with what is observed for
the 3D power spectrum (at $z=1$) in
Fig.~\ref{fig:inf_cl_kappa_conv}. When higher redshift sources
are included, we are sensitive to higher redshift, we expect the
departure from Gaussianity to be milder (non-linear evolution 
decreases with increasing redshift) and at smaller angular scales
(higher $\ell$). This corresponds to what is observed in the right
panel of Fig.~\ref{fig:inf_cl_kappa_conv}. 

\begin{figure}[t]
  \begin{center}
  \epsfig{angle=90,file=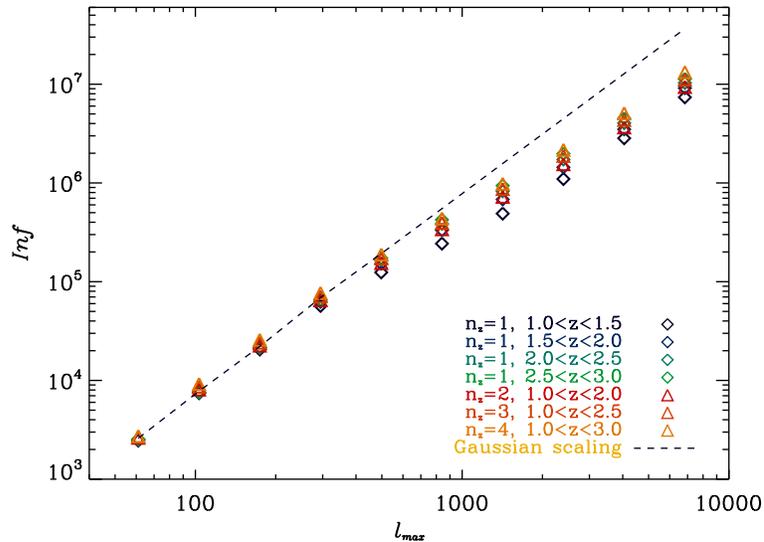,width=0.6\textwidth}
  \caption{\footnotesize Dimensionless cumulative information for the convergence
    cross-power spectra defined in Eq.~\ref{eq:inf_nodimcl}. Each
    color correspond to a different sub-set of source redshift band of size
    $n_z=$ 1, 2, 3 or 4. The dashed color line corresponds to the Gaussian prediction for the
   same number of modes. As in 3D, the effect of non-Gaussianity is clearly
   visible as a drop in the dimensionless information content when one
  enters the slightly non-linear regime. The scaling becomes Gaussian
  again in the fully non-linear regime. As expected, since non-linearities increase with redshift, the effect is more severe
  the lower the source redshift is.}
    \label{fig:inf_cl_kappa}
  \end{center}
\end{figure}

This is illustrated furthermore in Fig.~\ref{fig:inf_cl_kappa} where
we consider other source redshift distribution. Both the agreement on
large scale with the Gaussian prediction and the saturation effect shift to higher
$\ell$ with increasing source redshift are clearly visible. So is the
decreasing of the saturation effect as the source redshift increases
and when tomography is included. As for the 3D case, the information
increases in the fully non-linear regime, at a substantially lower value
than Gaussian. Note
however that the saturation effect is less dramatic than for the 3D
case since the projection inherent to lensing introduces an extra
Gaussianization.

\section{Discussion}
\label{sec:disc}

\begin{figure}[tbp]
  \begin{center}
  \epsfig{angle=90,file=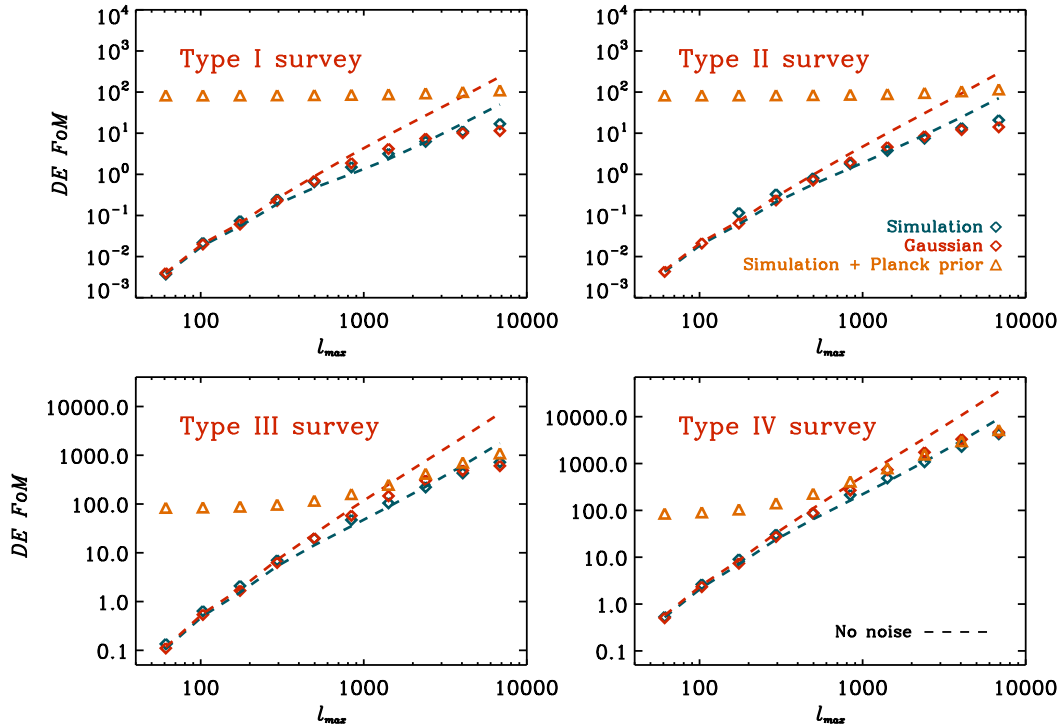,width=0.8\textwidth}
  \caption{\footnotesize Dark Energy Figure of Merit as a function of scales for 4 various
    surveys whose parameters are given in Tab.~\ref{Tab:survey}. The
    blue diamonds correspond to the predictions using non-Gaussian
    covariance matrix and a shot noise contribution. The red diamonds
    correspond to the Gaussian approximation to the Fisher matrix with
    shot noise. The orange diamonds correspond to the non-Gaussian case with the prior
    expected from the Planck satellite. The blue and red dashed curves
    correspond respectively to the non-Gaussian and Gaussian cases,
    without shot noise. Whereas we can see that the inclusion of non-Gaussian error
    bars is important for a perfect noiseless experiment (blue dashed
    curve) as compared to the perfect noiseless Gaussian errors (red
    dashed curve), it is less critical when adding the shot noise coming from the dispersion of
    intrinsic ellipticities (comparison between blue and red diamonds).}
    \label{fig:fom_surveys}
  \end{center}
\end{figure} 

Now that we highlighted above the dimensionless cumulative information
for some perfect idealized survey of angular area the corresponding to
the size of
our simulation, we discuss its implications for current and future optical
surveys. In particular, we quantify the consequences of the 
information saturation effects discussed above on the cosmological
information content. We focus on DE that we parametrize by an
evolving equation of state $w(a)=w_0+(1-a)w_a$. We use as our main
statistic the DE figure of merit (FoM) \cite{Albrecht:2006um}, that we
define as the area of the 95\% contour ellipse in the $w_0-w_a$ plane.
Following the definition of the Fisher matrix, $\bfF$, in
Eq.~\ref{eq:fisher_def}, if we consider the following observable
\be
\cP_{a=z_i(z_i-1)/2+z_j, \ell} = \tilde n_{z_i} \tilde n_{z_j} C_{\ell}^{\kappa\ z_iz_j}\,, \quad (i\ge j)
\ee
and a set of six cosmological parameters $\alpha_{\mu}$. The Fisher
information matrix writes as
\be
\bfF_{\mu \nu} = \sum_{\ell=2}^{\ell_{\rm max}}{\sum_{ab}{\partial{\cP_{a}}\over \partial{\alpha_{\mu}} } [ {\bf C}^{-1} ]_{ab}{\partial{\cP_{b}}
            \over \partial{\alpha_{\nu}}  }}\,.
\ee
With this notation, $\sigma(\alpha_\mu) = \sqrt{\bfF^{-1}_{\mu\mu}}$.

We will still make use of the 300 simulations introduced in
Sec.~\ref{Sec:CS_inf} and customized for the diffuse 21cm lensing
\cite{Lu:2009} to discuss optical cosmic shear. We do so by re-weighting the previous
results obtained using the wide uniform redshift bins with the
relative weights appropriate coming from a realistic optical galaxy density
\bea
n(z) & \propto  & z^\alpha e^{-\left( {z/z_0} \right)^\beta} , \\
{\rm with \quad} \alpha  = 2 &\quad {\rm and}& \beta  = 1.5 .
\eea 
$\tilde n_{i}$ corresponds to the total number of galaxies in the
$i$th redshift bin. We also rescale the signal covariance matrix by the survey area
considered. Note that in this work, we will ignore the uncertainties in
the number density, however important they are
\cite{Benjamin:2007ys}. We will consider four surveys whose parameters
are given in Tab.~\ref{Tab:survey}. They roughly correspond
respectively to a current survey like
CFHTLS \footnote{\texttt{http://www.cfht.hawaii.edu/Science/CFHLS/}},
the soon on-line
DES \footnote{\texttt{https://www.darkenergysurvey.org/}}  survey, and
wide and deep space survey like
Euclid \footnote{\texttt{http://sci.esa.int/science-e/www/area/index.cfm?fareaid=102}}
or JDEM \footnote{\texttt{http://jdem.gsfc.nasa.gov/}}. We will
normalize $n(z)$ so that the galaxy density matches the one given on
the second line of Tab.~\ref{Tab:survey} and we will consider the shot
noise coming from the intrinsic ellipticities of objects. Note that our predictions here are
somewhat inaccurate due to the fact we are re-weighting our 21cm
simulations and also our ideal survey is somewhat suboptimal since we
consider only 4 redshift bins to perform tomography, which has been
shown to be slightly sub-optimal \cite{Ma:2005rc}. However, our
treatment is accurate enough to discuss the effects of
non-Gaussianities that are the focus of our study. 

As a reference point, we will also compute the Fisher matrix with a
Gaussian approximation to $\bfC^{-1}$. Doing so, we follow the
formalism laid out for example in \cite{Ma:2005rc}.  In the Gaussian
case, assuming a shot noise level and simple Gaussian sample variance,
the covariance matrix is defined as  
\be
\bfC_{ab\ \ell} = \tilde n_{i} \tilde n_{j} \tilde n_{k} \tilde n_{l}
\left(C^{\rm tot\ ik}_{\ell} C^{\rm tot\ jl}_{\ell} + C^{\rm tot\ il}_{\ell}  C^{\rm tot\ jk}_{\ell}\right)\,,
\ee
where $a = i(i-1)/2+j$, $b= k(k-1)/2+l$ and where the total power spectrum is 
\be
C^{\rm tot\ ij}_{\ell}=C_{\ell}^{\kappa\ ij} + \delta_{ij} {\gamma_{\rm int}^2  \over {\tilde n}_i} \,,
\ee
where $\gamma_{\rm int}$ is the rms shear error per galaxy per component contributed by intrinsic ellipticity and measurement error.

Fig.~\ref{fig:fom_surveys} compares the Gaussian and
non-Gaussian cases for the 4 different surveys whose parameters are
defined in Tab.~\ref{Tab:survey}. Going from the Type I survey to the
Type IV survey, we increase simultaneously the number of galaxies, the
depth, the number of source redshift bins and the survey area hence an
increase in FoM. As we previously did for the dimensionless
information content, we will now study the evolution of the FoM as we
increase the number of modes, and go from the linear regime to the
non-linear regime. This is a proxy to quantify the cosmological
interpretation in this survey as we increase angular sensitivity. The red dashed lines corresponds to the noise free
Gaussian approximation while the blue dash line corresponds to the
non-Gaussian noise free evaluation. As we can see when comparing these
two curves, the saturation effect discovered earlier in the cumulative information translates
naturally in the evolution of the FoM with $\ell_{max}$. In the case
of a noiseless survey, the difference in FoM at high angular
resolution can be as high as a factor of ~4, even when we consider
four tomographic bins. This effect is thus important and in stark
contrast with the scaling with $\ell_{max}$ usually assumed in the
literature (\eg \cite{Amara:2006kp}). However, when introducing
realistic levels of noise, the effect is somewhat mitigated as is visible when
comparing the red diamonds (gaussian approximation, with noise) to the
blue diamond (non-gaussian covariance matrix, with noise). It is still
however non-negligible since the ratio between the Gaussian and
non-Gaussian cases at higher $\ell_{max}$ becomes closer to 1.6. This
point constitutes the key result from our study. While potentially
very damaging to the ideal performances of weak gravitational surveys,
the effect of non-Gaussianity is tampered by the estimated level of
Gaussian noise expected for current and future surveys. The inclusion
of the Planck prior does not affect those conclusions at high $\ell$. 

Note also that despite the fact that our plot hint at the ability to
measure the convergence power spectra up to $\ell<10000$, in practice,
it will most likely be limited by theoretical uncertainties at
$\ell\leq 3000$, not the least by our inability to model the details
of the baryon physics \cite{White:2004kv,Zhan:2004wq,Rudd:2007zx}.
\begin{figure}[tbp]
  \begin{center}
  \epsfig{angle=90,file=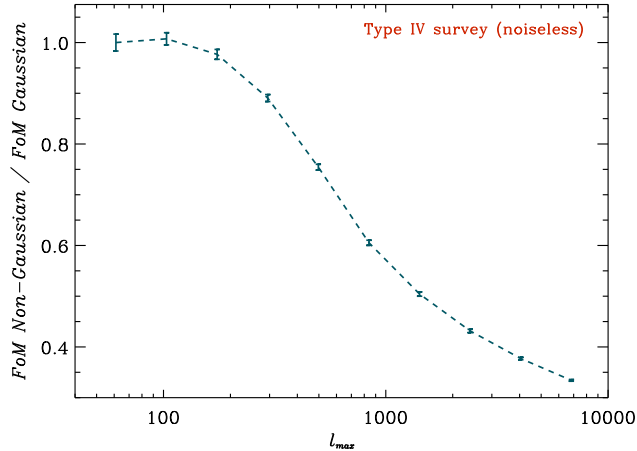,width=0.5\textwidth}
  \caption{\footnotesize Bootstrap errors on the ratio between the
    noiseless FoM curves for the Type IV survey. We are thus plotting
    the ratio between the red and blue dashed curve  in the lower
    right panel of Fig.~\ref{fig:fom_surveys}}     \label{fig:fom_error}
  \end{center}
\end{figure} 

\begin{figure}[t]
  \begin{center}
  \epsfig{angle=90,file=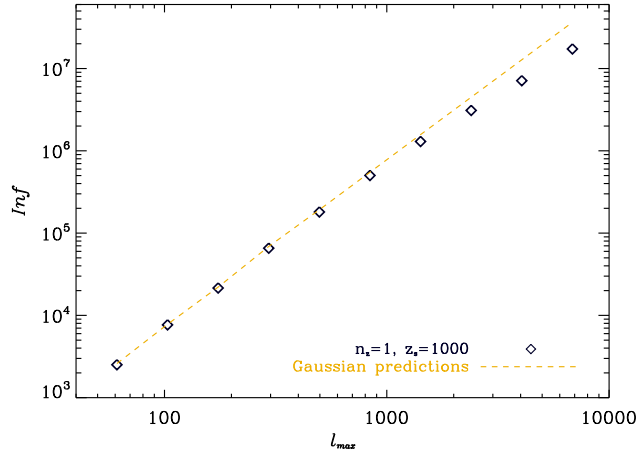,width=0.5\textwidth}
  \caption{\footnotesize Dimensionless cumulative information content
    as a function of maximum multipole for a source plane at
    $z=1000$. A mild saturation effect is still visible. Since the CMB lensing is sensitive mostly to the growth
    of structure at $z\simeq 2.5$, the saturation effect is milder and
    shifted to smaller angular scales (higher $\ell_{max}$).}
    \label{fig:inf_cmb}
  \end{center}
\end{figure} 

\begin{table}[b]
\begin{center}
\caption{Optical surveys considered}
\begin{ruledtabular}
\begin{tabular}{c|cccc}
Survey & I &  II & III & IV  \\
\hline
$Area$ (deg.$^2$) & 200 &  200  & 5 000 & 20 000\\
\hline
$n_{gal}$ (\#/arcmin.$^2$) & 30 & 30 & 50 & 100\\
\hline
$\langle\gamma\rangle$ & 0.25 & 0.25 & 0.25  &  0.25\\
\hline
\# redshift bins & 1 &  3  & 3 & 4 
\label{Tab:survey}
\end{tabular}
\end{ruledtabular}
\end{center}
\end{table}

Consistent conclusions were reached in the halo model based analytical approach
followed in \cite{Cooray:2000ry,Takada:2008fn}. 
We will discuss the
latter as being the most recent and refer to its discussion of the
earlier results of \cite{Cooray:2000ry}. Our results seem consistent
with their evaluation of the signal to noise ratio (somewhat
comparable to our dimensionless information). So does the effects on
parameters when considering realistic surveys although our factor of
~1.6 for FoM seems a bit higher. Note however that the comparison is
not direct since the set of parameters we consider is slightly
different. Whereas we consider a flat Universe with 6 parameters
including an evolving DE equation of state, they consider non-flat
models and allow the spectral index to run. We did not translate our
evaluation of the cross-power spectrum covariance matrices to real
space statistics and therefore we refer the reader to the discussion
of \cite{Takada:2008fn} for a comparison with real space evaluation of
this effect, as \eg  \cite{Semboloni:2006gc}. A more detailed
comparison between analytical estimates and numerical work would be
instructive and has to be performed but we leave it for further
work. So do we for the dependence of our results on cosmological
parameters, mostly $\sigma_8$, $n_s$ and $\Omega_m$
\cite{Eifler:2008gx}. Note that we also do not investigate the
so-called ``beat-coupling'' effect  \cite{Hamilton:2005dx,Takada:2008fn}. As we
understand it, it denotes the extra-mode coupling induced by the
finiteness of the observed volume, \ie such an effect would be
non-existent if the all-sky was considered. Since the exact form of
this coupling will depend on the exact mask of any given survey, we
decided to ignore this effect here. In practice, as it is usually
performed for example while analysing CMB data, we would start from a
mask description in real space and propagate the induced mode-coupling
in Fourier space throughout all our calculations
\cite{Padmanabhan:2002yv} with special care to the peculiarities of
non-Gaussian statistics. With their own prescription, \citet{Takada:2008fn} found that it does not affect
qualitatively the effect of the non-linear growth of structures.
 
The comparison with the numerical work of \cite{Hu:2000ax} is also not
obvious. First, a different cosmology with a higher $\sigma_8$ is
considered, which should enhance the non-linear effect. Second, they
consider 200 simulations and a standard tilling technique that should
give rise to an accuracy of at most at most 85\% on the errors using
the scaling formula of \citet{Takahashi:2009bq}. All those reasons
make a direct comparison a bit difficult. 

To quantify further the error on the errors issue, we evaluate the errors of the FoM
for the Type IV survey introduced before using a bootstrap method
discussed above. We display the results in Fig.~\ref{fig:fom_error} where we plotted the
relative difference between the Gaussian and non-Gaussian FoM as a
function of $\ell_{max}$. Due to the strong convergence properties of
our covariance estimation technique, we quantify the error on the
errors to be around 25\% on large scale and sub-percent on the smaller
scales we consider. This fact certainly constitutes an improvement over
previous results in the literature and is consistent with our
estimate of a 1/$\sqrt{\cns n_{box}}$ convergence rate. If we use the
scaling of 12/$\sqrt{\cns}$ measured  by \citet{Takahashi:2009bq}, we claim an overal
uncertainty due to the limited number of simulations to be around 14\%.

Besides the optical observations of weak-gravitational lensing, the
cosmic microwave background (CMB) constitutes another source plane
where lensing can be observed (see \cite{Lewis:2006fu} for a
review). As both the resolution and the sensitivity of detectors
improve, it can now be measured and it defines the next frontier for
the CMB temperature and polarization measurement \cite{Smith:2008an}.
Using cross correlation between WMAP data and other tracers of large scale
structures to increase the signal to noise, a detection of
gravitational lensing in the CMB temperature has been achieved with
marginal significance, i.e. around 2.4 $\sigma$
\cite{Hirata:2004rp,Smith:2007rg,Hirata:2008cb}. A direct detection in
temperature is expected to be achieved soon with high significance thanks to
on-going high angular resolution temperature surveys (ACT, SPT,
Planck). It is thus interesting to evaluate the effects of the
non-linear growth of structures on the CMB lensing signal. To do so, we compute the
cumulative information as a function of maximum angular scale
$\ell_{max}$, when considering one redshift source plane at
$z=1000$. The result is displayed in Fig.~\ref{fig:inf_cmb}. Interestingly, the saturation effect
is still visible, and not surprisingly, since the CMB lensing kernel
peaks around z$\simeq 2.5$, the effect is smaller and shifted to
smaller scales. In practice however, given that the CMB reconstruction is
most likely going to be limited by secondary anisotropies (kinetic SZ
in particular and patchy reionization) around a few $\ell\simeq 3000$,
it is unlikely that this effect of non-linear growth will be a critical effect.

To conclude, non-Gaussian effects are potentially very important for weak-gravitational
surveys and might alter significantly the forecasts done so far. When
considering realistic noise estimate for the coming optical surveys,
the impact of non-Gaussian error bars is much milder. However, an
interesting indirect consequences of our study is the sub-Gaussian
scaling of information in the fully non-linear regime (this was also
observed in the 3D power spectrum \cite{Rimes:2005dz}). This fact highlights
the great information gain that can be make by studying this regime, both using
2 point statistics but also higher order more specific to non-Gaussian
effects. Nevertheless, it will require more theoretical insights.

\acknowledgments{We thank the participants of the CITA/CIFAR workshop
  ``Upcoming Lensing surveys: Beyond The obvious'' for stimulating
  discussions that improved this work. We also thank Dick Bond and
  Martin White for stimulating remarks. All computations were
  performed on the Canada Foundation for Innovation funded CITA
  Sunnyvale cluster.} 

\bibliography{general_bib}

\end{document}